\documentclass[conference]{IEEEtran}
\usepackage{pifont}
\newcommand{\checkmark}{\ding{51}}
\newcommand{\xmark}{\ding{55}}
\IEEEoverridecommandlockouts
\usepackage{cite}
\usepackage{algorithmic}
\usepackage{algorithm}
\usepackage{graphicx}
\usepackage{textcomp}
\usepackage{amsmath,amssymb,amsfonts}
\usepackage{stfloats}
\usepackage{multirow}
\usepackage{xcolor}
\usepackage{url}
\usepackage{placeins}
\usepackage{enumitem}
\usepackage{pgfplots}
\usepackage{caption}
\usepgfplotslibrary{statistics}
\pgfplotsset{compat=1.18}  % or a version suitable to your system
\usetikzlibrary{plotmarks}
\usetikzlibrary{patterns}
\usepackage{pgfplotstable}
\usepackage{filecontents}
\usepackage{adjustbox}
\usepackage{booktabs}
\usepackage{diagbox}
\usepackage{geometry}
\usepackage{afterpage}
\usepackage{subcaption}
\usepackage{geometry}
\def\ourmethod{GETA}

\def\BibTeX{{\rm B\kern-.05em{\sc i\kern-.025em b}\kern-.08em
    T\kern-.1667em\lower.7ex\hbox{E}\kern-.125emX}}
\begin{document}
\bstctlcite{IEEEexample:BSTcontrol}

\title{GETA: Generalized Encrypted Traffic Analysis

}

% \and
% \and
\author{\IEEEauthorblockN{Ransika Gunasekara, Rahat Masood, and Salil Kanhere}
\IEEEauthorblockA{\textit{University of New South Wales (UNSW)} \\
Sydney, Australia}
}

\maketitle

\begin{abstract}
 Traditional traffic analysis is being fundamentally challenged by the rapid adoption of encryption, tunnelling, and privacy-preserving protocols, which increasingly obscure packet payloads and limit the usefulness of Deep Packet Inspection (DPI).  Although machine learning has advanced encrypted traffic analysis, existing approaches often remain tied to protocol-specific header features, depend on large labelled datasets, and degrade when deployed across heterogeneous network environments. We present GETA, a protocol-agnostic framework for encrypted traffic analysis that models network flows as multivariate time series using only traffic metadata, thereby avoiding reliance on packet payloads or header semantics. GETA combines meta-learning, embedding refinement, and self-attention to support few-shot adaptation to previously unseen domains with minimal labelled data. Across nine public datasets spanning application identification, VPN traffic classification, IoT device fingerprinting, and attack detection, GETA consistently outperforms state-of-the-art baselines. These results show that GETA offers a practical and generalisable foundation for robust traffic analysis in modern encrypted networks.
\end{abstract}


\section{Introduction}

The widespread adoption of encryption protocols such as TLS, QUIC, and IPsec enhances the confidentiality and integrity of network communications but limits the effectiveness of traditional monitoring methods, such as deep packet inspection (DPI), that rely on payload inspection. Consequently, conventional traffic analysis struggles to characterize and classify encrypted flows. In contrast, machine learning (ML) approaches exploit observable flow features such as packet sizes, inter-arrival times, and directions to analyze traffic without accessing payloads~\cite{timeseries_ETA1,timeseries_ETA2}, enabling scalable and privacy-preserving visibility into encrypted networks.

Encrypted traffic analysis (ETA) has emerged as a key paradigm for understanding network behavior and detecting malicious activity through inference from traffic patterns, flow dynamics, and statistical characteristics ~\cite{Papadogiannaki2021}. With over 90\% of web traffic now encrypted~\cite{google_transparency_https}, the ability to analyze traffic flows without inspecting packet contents has become essential for both research and operational network security.

Advances in ML and deep learning (DL) have enhanced ETA by enabling automated extraction of complex, non-linear patterns in encrypted traffic, improving classification and behavioral modeling~\cite{Yang2024}. However, several approaches still rely on packet header information~\cite{UMVD,ETBERT}, a major limitation in scenarios involving VPNs, tunnels, or privacy-preserving networks where headers are minimal, encrypted, or entirely obfuscated,  leading to degraded accuracy and robustness across diverse environments~\cite{Appsniffer}.

A key challenge for ML-based ETA is its reliance on large labeled datasets. While supervised learning performs well in controlled settings, it demands extensive data collection and manual labeling, limiting scalability in real-world deployments~\cite{Shen2023}. Changes in traffic patterns, device usage, or encryption protocols often demand retraining, which is time-consuming and resource-intensive. %in dynamic operational settings.
This reliance on static datasets also weakens generalizability. Models trained in one environment frequently fail in another because they overfit to environment-specific characteristics instead of learning robust, transferable traffic patterns~\cite{wu2024encrypted, malekghaini2023deep}. Recent studies show sharp performance drops under domain shifts~\cite{ETBERT,Appsniffer}, highlighting the brittleness of current ETA solutions.
%Without efficient adaptation mechanisms, ETA remains brittle and unreliable in real-world deployments.

To address these challenges, we present \textbf{GETA: Generalized Encrypted Traffic Analysis}, a protocol-agnostic framework that models encrypted traffic as multivariate time series using only consistently observable traffic metadata. Rather than relying on packet payloads or header semantics, GETA jointly models packet size, inter-arrival time, and packet direction as correlated temporal signals. This design enables GETA to capture richer flow-level dependencies while remaining applicable across diverse encrypted and tunneled network environments. To further improve generalization, GETA integrates meta-learning with embedding refinement and self-attention, allowing the model to adapt to unseen network domains using only a small number of labelled examples. This makes GETA particularly suitable for realistic deployments where traffic distributions evolve and labelled data are scarce. This work makes the following contributions:

\begin{itemize}[leftmargin=0pt]
   \item \textbf{A protocol-agnostic representation for encrypted traffic analysis.}
We introduce a metadata-driven multivariate time-series representation for ETA that avoids reliance on packet payloads, protocol headers, or application-specific semantics. By jointly modeling packet size, inter-arrival time, and packet direction, GETA captures temporal and cross-feature dependencies that are often missed by univariate or header-dependent approaches.

      \item \textbf{Few-shot generalization through meta-learning and embedding refinement.}
    We present a meta-learning-based methodology for ETA under diverse network conditions. We propose a novel modification to time-series models to handle multivariate traffic metadata.
    \ourmethod\ incorporates a meta-learning framework that enables rapid adaptation using only a few labeled samples. A key innovation is the embedding enhancement module, which refines the initial traffic embeddings to capture more nuanced representations. In addition, we apply self-attention over the support set to inject task-specific context into the embeddings, enabling the model to adapt more effectively to each classification task. Together, these components produce more discriminative and generalizable representations, boosting performance across diverse network conditions without requiring task-specific retraining.

    \item \textbf{Strong generalization and superior performance across diverse domains and tasks:}
    GETA demonstrates robust few-shot performance under domain shifts, reflecting realistic deployment scenarios with limited labeled data and evolving network conditions. We meta-train on one dataset and evaluate on a different dataset across nine public datasets, spanning application identification (including VPN traffic), IoT device classification, and attack detection. GETA consistently outperforms existing baselines compared to the next-best model in diverse tasks.
\end{itemize}
The remainder of the paper is organized as follows: Section II discusses related work, Section III describes the proposed methodology, Section IV presents evaluation results across multiple datasets, and Section V concludes with a discussion.

\section{Related Work}

We organize the related work into three areas aligned with our methodology. First, we review deep learning methods for ETA, followed by Few-Shot/Zero-Shot Learning approaches that address data scarcity. Next, we examine Time Series Modeling, which motivates our use of structured traffic metadata. %This structure highlights how our work bridges gaps across these domains.

\subsection{Encrypted Traffic Analysis (ETA)}

ETA seeks to infer application, device, or behavioral information from encrypted network flows without accessing payloads. Recent transformer- and LLM-inspired methods operate directly on raw packet bytes ~\cite{ETBERT,flow_mae,netmamba}. For instance, ET-BERT ~\cite{ETBERT} applies a BERT-style model to Ethernet packet bytes, achieving high accuracy but requiring extensive pretraining and showing poor generalization in constrained settings such as VPNs or non-TCP traffic ~\cite{Appsniffer}. These limitations stem from a reliance on byte-level patterns and plaintext headers, which may be obfuscated or unavailable in practice. Furthermore, ~\cite{wickramasinghe2025sok} shows that byte-based models frequently overfit to pseudorandom fields such as Sequence/Ack numbers, learning spurious correlations rather than true traffic semantics. Metadata-based ETA methods, which predate byte-based deep learning,  instead rely on flow-level features such as packet size, direction, and timing ~\cite{koumar2023,doh_detection}. While these methods offer greater interpretability and robustness to encryption or protocol variation, they traditionally require large labeled datasets and remain domain-specific~\cite{Var-CNN,Deng2025,Appsniffer}. \textit{As byte-based methods remain vulnerable to obfuscation and domain shifts,  revisiting metadata-based analysis with modern learning paradigms is essential for building generalizable and data-efficient ETA systems.}

\subsection{Few-Shot/Zero-Shot Learning for ETA}

Few-shot learning addresses data scarcity in traffic classification amid rapid shifts in device types and application behaviors. Zhao et al.~\cite{Zhao2022} use the first few bytes of initial packets for IoT traffic classification, achieving good results in TCP/UDP settings, but as our RQ3 experiments show, performance degrades sharply when header information is minimal. Others~\cite{Yang2023,MetaRocket} propose few-shot learning methods that use directed packet-length sequences with Model-Agnostic Meta-Learning (MAML)~\cite{Finn2017}, improving adaptation and reducing dependence on raw bytes, but capture only packet size and direction while neglecting temporal structure. More recently, ZEST~\cite{Zest} explores zero-shot IoT device classification to identify unseen devices, showing promise but remaining limited to one dataset and requiring large labeled data for seen classes. 

\textit{In contrast, our work integrates few-shot learning with a transformer-based multi-variate time-series model, enabling robust generalization to unseen traffic classes with minimal labeled data and strong performance across heterogeneous classification tasks such as application, device, and attack identification.}

\subsection{Time Series Modeling for Network Traffic}

Several ETA methods model encrypted traffic as a time series for classification~\cite{koumar2023,timeseries_ETA1, timeseries_ETA2}, but they often rely on single-variable representations (e.g., packet size) and require manually crafted features. This process is time-consuming, error-prone, and often fails to capture complex temporal dependencies among flow attributes ~\cite{ETBERT,timeseries_ETA1, timeseries_ETA2}.

Time series models address this by treating flows as sequences of structured metadata (e.g., packet size, inter-arrival time, direction), providing invariance to encryption protocols and payload content. MetaRockETC~\cite{MetaRocket} adopts a related approach using MultiRocket~\cite{multirocket}. Recent advances in transformer-based architectures have further improved time-series analysis ~\cite{transformerstimesries,time_series_transformers1,time_series_transformers2,UniTS}. Notably, UniTS ~\cite{UniTS} presents a unified framework that performs well across diverse tasks and supports multivariate time series inputs. However, UniTS is not designed for few-shot learning and lacks mechanisms for rapid adaptation or cross-domain generalization.

\textit{Our method extends UniTS beyond general-purpose time-series modeling by incorporating meta-learning for few-shot ETA. We tailor the architecture to network-specific multivariate inputs and evaluate it across heterogeneous downstream tasks, demonstrating adaptability for real-world security deployments.}

% \subsection{Meta-Learning in ETA}

% Meta-learning, particularly gradient-based methods like MAML ~\cite{Finn2017}, has shown strong potential in domains where labeled data is limited and tasks are diverse. In network traffic analysis, meta-learning allows models to quickly adapt to new environments, traffic types, or behavioral patterns with minimal supervision. However, most applications in network security are limited to synthetic datasets or narrowly defined tasks. Few studies investigate multivariate time series models trained within a meta-learning framework, or systematically evaluate generalization across real-world tasks such as app identification, device fingerprinting, and attack detection. Some ETA methods ~\cite{Zheng2020, Yang2023}  incorporate meta-learning to support few-shot adaptation, but their evaluation on cross-domain generalization remains limited. \textit{Our work addresses this gap by integrating meta-learning with a modified time-series transformer, enabling robust few-shot classification while generalizing across heterogeneous domains. By jointly leveraging task-agnostic learning and structured temporal modeling, two properties rarely combined in prior work, GETA advances the applicability of meta-learning to practical and security-critical scenarios.}

\section{Methodology}

\subsection{Overview and Notation}
\label{sec:notation_table}

\begin{figure*}[t]
\centering
\resizebox{0.9\textwidth}{!}{%
\includegraphics[trim=0.4cm 0cm 4cm 0cm, clip]{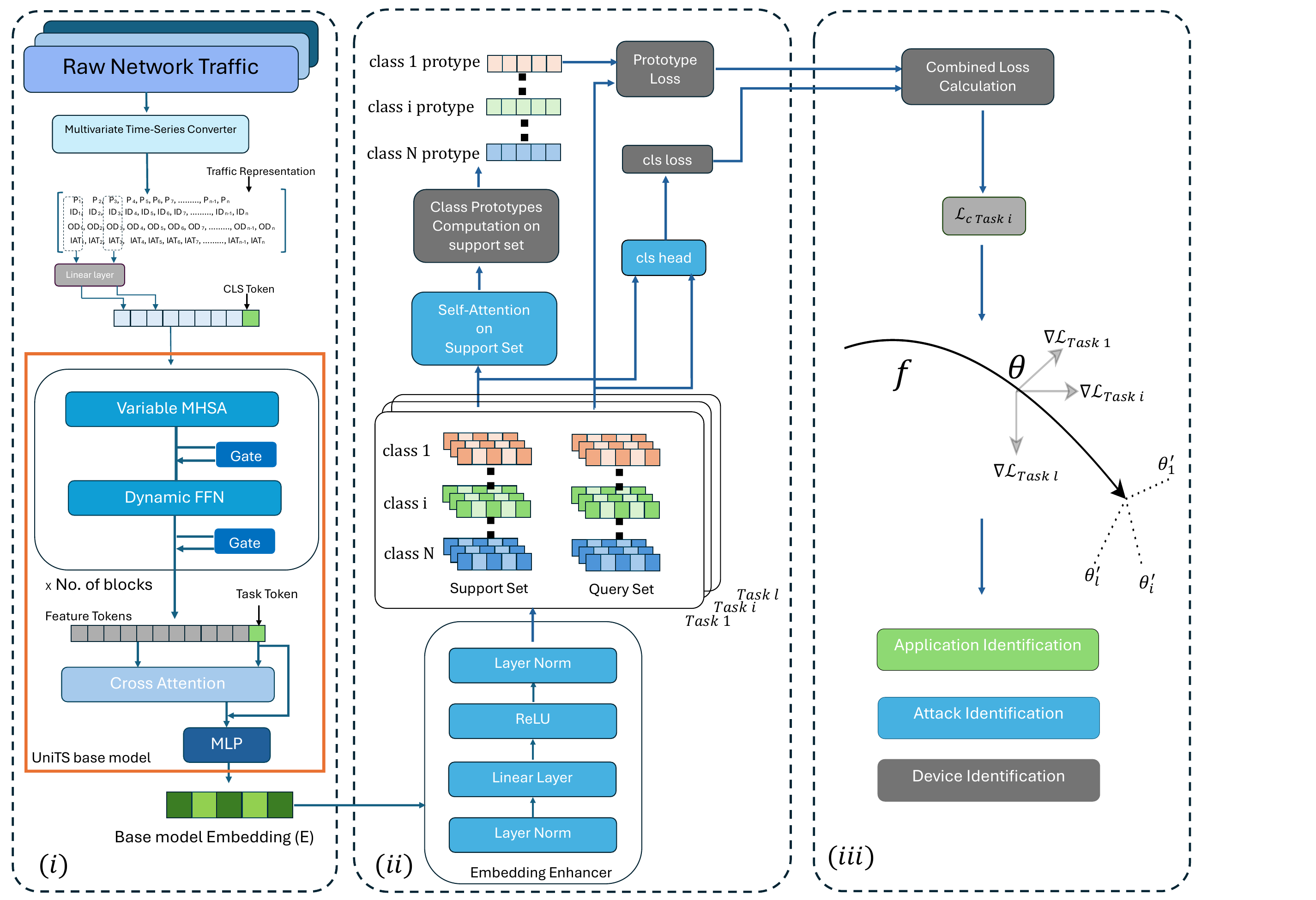}
}
\caption{Overall Methodology of \ourmethod. \textbf{(i)} shows the traffic representation as a multivariate time series and embedding generation via the base model. \textbf{(ii)} shows the embedding enhancement from the base model and prototype enhancement with the use of self-attention and the calculation of prototype loss and cls loss. \textbf{(iii)} shows the meta-training stage for optimal parameter initialization and the fine-tuning to the downstream tasks.}
\label{fig:methodology}
\end{figure*}

\ourmethod\ leverages meta-learning to advance ETA under few-shot learning settings, enabling adaptation to downstream tasks with minimal labeled data. Its objective is to learn an embedding function and a classification mechanism that, after a few inner-loop updates on a small support set, generalize effectively to query samples from the same task. Our approach, shown in Figure~\ref{fig:methodology}, consists of five stages: traffic representation, embedding generation, embedding enhancement, self-attention-based prototype creation, and meta-learning optimization. Each stage supports this objective: traffic representation and base embeddings capture task-agnostic features; the embedding enhancer and prototype refinement improve discriminability in few-shot decision boundaries; and meta-learning optimises initialisation for fast adaptation.

Table~\ref{tab:notation} provides a complete reference for all notation and dimensions used throughout this section.

\begin{table}[h]
\caption{Notation and dimensions. }
\label{tab:notation}
\centering
\begin{tabular}{@{}l@{}}
\toprule
$s$, $v$, $d$: Seq.\ len($s$=512), \#variables ($v{=}4$), hidden dim ($d{=}64$) \\
$d_E$: Embedding dim.\ ($d_E = v \cdot d = 256$) \\
$X$: Input (after patch embed.); $\mathbb{R}^{s \times v \times d}$. \\
$E$, $E'$: Base and enhanced embedding; $\mathbb{R}^{d_E}$. \\
$E'_k$, $E'_q$: Enhanced support (class $k$) and query; $\mathbb{R}^{d_E}$. \\
$A_k$: Refined support embeddings; $\mathbb{R}^{|D^s_k| \times d_E}$. \\
$C_k$, $C'_k$: Prototype and ProtoAdaptation output; $\mathbb{R}^{d_E}$. \\
$\tau$, $\lambda$: Temperature, loss weight; learnable / hyperparam. \\
$D^s$, $D^q$, $D^s_k$: Support, query, support for class $k$; sets. \\
\bottomrule
\end{tabular}
\vspace{-10 pt}
\end{table}

\subsection{Traffic Representation}

We represent network traffic as a multivariate time series to capture the joint temporal dynamics of packet size, direction, and timing. This protocol-agnostic representation relies on observable features that remain available under encryption and VPNs. We use four variables: packet size ($P_t$), incoming direction ($ID_t$), outgoing direction ($OD_t$), and inter-arrival time ($IAT_t$). The traffic sequence at time step $t$ is:
{\setlength{\abovedisplayskip}{0pt}
\setlength{\belowdisplayskip}{-0pt} 
\begin{equation*}
    X_t = (P_t, ID_t, OD_t, IAT_t),  X = \{X_1, X_2, \dots, X_T\}
\end{equation*}}

Each tuple represents the four features at time step $t$, where $X$ is the input sequence of length $T$. For example, a traffic sequence of length $T=3$ might look like: $
X = \{(512, -1, 0, 0.0),\ (128, 0, 1, 0.1),\ (64, -1, 0, 0.8)\}$. 

We represent traffic direction using two binary indicators $(ID_t, OD_t)$ rather than a single binary variable. Empirically, this increased classification accuracy by up to 1\% compared to using a single variable, likely due to clearer directional semantics, improved gradient flow, and reduced feature ambiguity during training.

\subsection{Base Model Embedding Generation}

To extract meaningful representations from the multivariate traffic sequence, we adopt a modified transformer-based architecture based on UniTS~\cite{UniTS}, shown in Fig~\ref{fig:methodology} part(i). UniTS is chosen for its ability to model multivariate time-series data. The input tensor $X \in \mathbb{R}^{s \times v \times d}$ (see Table~\ref{tab:notation} for dimensions) undergoes variable-wise multi-head self-attention to capture feature interactions, with attention weights computed between variables and averaged across the time dimension for efficiency. The variable-attentive representations are modulated by learnable gating and passed through a Dynamic Feed Forward Network with temporal convolution to capture local sequential patterns. A learned task token then queries all feature tokens via cross-attention (scaled dot-product, task token as query), followed by MLP refinement, producing the final embedding $E \in \mathbb{R}^{d_E}$:
{\setlength{\abovedisplayskip}{0pt}
\setlength{\belowdisplayskip}{-5pt} 
\begin{align}
    E = \text{\textit{MLP}}(&\text{\textit{CrossAttention}}(\text{\textit{task token}}, \nonumber \\
                  &\text{\textit{feature tokens}}, \text{\textit{feature tokens}}))
\end{align}
}

\subsection{Embedding Enhancer Module}

We introduce an embedding enhancement module to refine embeddings before prototype computation and classification. This is motivated by the need to reduce noise and improve discriminability in the embedding space when only a few support samples per class are available. The module stabilizes gradients and keeps embeddings well-scaled via layer normalization, a linear projection with ReLU, and a final layer norm. Given input embedding $E \in \mathbb{R}^{d_E}$, the enhanced embedding $E'$ (same dimension) is:
\setlength{\abovedisplayskip}{0pt}
\setlength{\belowdisplayskip}{0pt}
\begin{equation}
\begin{split}
    E' &= \text{LayerNorm}(\text{ReLU}(W E + b)),\\
    &\quad W \in \mathbb{R}^{d_E \times d_E},\; b \in \mathbb{R}^{d_E}.
\end{split}
\end{equation}
In our implementation, we additionally apply dropout and an optional feature-wise scale and bias (calibration) for stability. $E'_k$ and $E'_q$ denote enhanced embeddings of support samples for class $k$ and of the query sample, respectively.

\subsection{Prototype Based Loss}

Based on ~\cite{snell2017prototypical}, this stage aims to form class representatives that are robust to support-set noise and transfer across tasks. For each class $k$, we first refine the support embeddings $E'_k$ with multi-head self-attention (scaled dot-product, 4 heads): the same set of vectors is used as query, key, and value, yielding refined embeddings $A_k$. The class prototype is their mean. Formally, for class $k$ with support set $D^s_k$:
{\setlength{\abovedisplayskip}{0pt}
\setlength{\belowdisplayskip}{0pt}
\begin{equation}
\begin{aligned}
A_k &= \text{MultiHeadSelfAttn}(E'_k, E'_k, E'_k) 
      \in \mathbb{R}^{|D^s_k| \times d_E},\\
C_k &= \frac{1}{|D^s_k|} \sum_{(x_i, y_i) \in D^s_k} A_i 
      \in \mathbb{R}^{d_E}.
\end{aligned}
\end{equation}}

\textbf{ProtoAdaptation} is a two-layer MLP that maps prototypes to a space better suited for cross-task comparison: $C'_k = \text{ProtoAdaptation}(C_k)$ with
$\text{ProtoAdaptation}(z) = W_2 \,\text{ReLU}(W_1 z + b_1) + b_2$, where $W_1, W_2 \in \mathbb{R}^{d_E \times d_E}$ and $b_1, b_2 \in \mathbb{R}^{d_E}$ are learnable. For the query sample $x_q$ with enhanced embedding $E'_q$, we compute temperature-scaled squared Euclidean distances:
{\setlength{\abovedisplayskip}{0pt}
\setlength{\belowdisplayskip}{0pt}
\begin{equation} d_k = \frac{|E'_q - C'_k|_2^2}{\tau} \end{equation}}
where $\tau$ is a learnable temperature parameter. Classification probabilities are obtained via softmax over negative distances:
{\setlength{\abovedisplayskip}{0pt}
\setlength{\belowdisplayskip}{0pt}
\begin{equation} p(y = k \mid x_q) = \frac{e^{-d_k}}{\sum_j e^{-d_j}} \end{equation}}
The prototype loss is the cross-entropy between predicted and ground-truth labels, serving as one branch of our dual-pathway loss function (Section~\ref{sec:combined_loss}).

\subsection{Combined Loss Function}
\label{sec:combined_loss}

To combine the strengths of prototype-based learning and conventional classification, we use a dual-pathway approach with a combined loss function. This provides richer gradient signals during adaptation while retaining a simple prototype-based decision rule at test time:
{\setlength{\abovedisplayskip}{0pt}
\setlength{\belowdisplayskip}{0pt}
\begin{equation}
    \small
    \mathcal{L}_{combined} = (1 - \lambda) \cdot \mathcal{L}_{proto} + \frac{\lambda}{2} \cdot (\mathcal{L}_{cls\_query} + \mathcal{L}_{cls\_support})
    \label{eq:combined_loss}
\end{equation}}

where $\lambda$ is a weighting hyperparameter that balances both pathways. The prototypical network loss $\mathcal{L}_{proto}$ enforces a well-structured embedding space with same-class examples clustered together. The classification losses $\mathcal{L}_{cls\_query}$ and $\mathcal{L}_{cls\_support}$ ensure that these embeddings remain linearly separated through direct mapping. This dual formulation provides rich gradient information and acts as an implicit regularizer, preventing overfitting to a single pathway and fostering more robust, generalizable features.

We employ Model-Agnostic Meta-Learning (MAML) to optimize initialization parameters $\theta$ for rapid adaptation to new tasks (Algorithm~\ref{alg:meta_learning}). This is motivated by the objective of finding an initialization from which a few gradient steps on the support set yield good query-set performance. For each task $\mathcal{T}_i$, we perform inner-loop adaptation on the support set:
{\setlength{\abovedisplayskip}{0pt}
\setlength{\belowdisplayskip}{0pt}
\begin{equation}
    \theta'_i = \theta - \alpha \nabla_\theta \mathcal{L}(D^s_{T_i}, f_\theta)
\end{equation}}

The meta-update aggregates gradients across tasks using the query sets:
{\setlength{\abovedisplayskip}{0pt}
\setlength{\belowdisplayskip}{0pt}
\begin{equation}
    \theta \leftarrow \theta - \beta \sum_{i} \nabla_\theta \mathcal{L}(D^q_{T_i}, f_{\theta'_i})
\end{equation}}
where $\alpha$ is the inner-loop (task) learning rate and $\beta$ is the meta (outer) learning rate. 

We use $N_{adapt}=3$ inner-loop adaptation steps during meta-training and $20$ at evaluation. On the first step, the $\mathcal{L}_{combined}$ is optimised, and on subsequent steps, only $\mathcal{L}_{proto}$. Training settings: $\alpha=0.001$, $\beta=0.0001$, $\lambda=0.5$; patch length and stride 4; encoder depth 3; dropout 0.3 and batch size 32.

\subsection{Progressive Fine-Tuning Strategy}

After meta-training, we fine-tune on downstream tasks (e.g., 5-way 5-shot) using a progressive adaptation strategy:

\begin{enumerate}
    \item \textbf{Initial Comprehensive Adaptation}: Optimize the combined loss function, incorporating both prototypical and classification losses for rich gradient information.
    
    \item \textbf{Subsequent Focused Adaptation}: Use only prototypical loss to improve efficiency and reduce overfitting to the small support set.
    
    \item \textbf{Final Evaluation}: Assess performance on the query set using prototype-based classification, ensuring robust feature learning through the learned prototypes.

\end{enumerate}

This strategy enables efficient knowledge transfer while remaining robust to limited labelled data, making it well-suited to real-world ETA scenarios.
\subsection{Reproducibility}
All hyperparameters and dimensional settings are summarized in Table~\ref{tab:notation}; evaluation protocols are specified in Section~\ref{sec:evaluation}. Source code and a self-contained reproducibility bundle are available at \cite{GETAcode}.

\begin{algorithm}[t]
\small
\caption{Model Agnostic Meta-Learning with Combined Loss}
\begin{algorithmic}[1]
\REQUIRE Dataset $\mathcal{D}$, embedding function $f$, hyperparameters $\alpha$, $\beta$, $\lambda$, number of adaptation steps $N_{adapt}$
\ENSURE Optimal parameters $\theta^*$
\STATE Create tasks $\{\mathcal{T}_1, ..., \mathcal{T}_l\}$ from $\mathcal{D}$, split into support $D^s$ and query $D^q$ sets
\STATE Initialize model parameters $\theta$
\FOR{each epoch}
  \STATE Sample batch of tasks $B = \{\mathcal{T}_i\}$
  \FOR{each task $\mathcal{T}_i \in B$}
    \STATE \textbf{Inner loop:} Clone $\theta'_i \leftarrow \theta$
    \FOR{adaptation steps}
      \IF{first step}
        \STATE $\mathcal{L} = (1-\lambda)\mathcal{L}_{proto} + \frac{\lambda}{2}(\mathcal{L}_{cls\_query} + \mathcal{L}_{cls\_support})$
      \ELSE
        \STATE $\mathcal{L} = \mathcal{L}_{proto}$
      \ENDIF
      \STATE $\theta'_i \leftarrow \theta'_i - \alpha \nabla_{\theta'_i} \mathcal{L}(D^s_{\mathcal{T}_i}, f_{\theta'_i})$
    \ENDFOR
    \STATE \textbf{Outer loop:} Evaluate on $D^q_{\mathcal{T}_i}$ with stochastic loss selection
    \STATE Accumulate gradients: $G \leftarrow G + \nabla_{\theta} \mathcal{L}(D^q_{\mathcal{T}_i}, f_{\theta'_i})$
  \ENDFOR
  \STATE \textbf{Meta-update:} $\theta \leftarrow \theta - \beta G$
\ENDFOR
\RETURN $\theta^*$
\end{algorithmic}
\label{alg:meta_learning}
\end{algorithm}

\section{Evaluation}
\label{sec:evaluation}

In this section, we evaluate \ourmethod’s performance in various ETA scenarios and investigate its generalization capabilities under limited labeled data constraints. Our experiments are designed to answer the following research questions:

\begin{itemize}[leftmargin=0pt, label={}]
    \item \textbf{RQ1:} How well does \ourmethod\ generalize across heterogeneous datasets and network environments, including device, application, and attack classification tasks?
    \item \textbf{RQ2:} How robust is \ourmethod under diverse N-way K-shot learning settings?
    \item \textbf{RQ3:} Can \ourmethod's metadata-based approach maintain strong performance in VPN-encrypted environments where header information is minimal or unavailable? 
    \item \textbf{RQ4:} How does each main component of \ourmethod\ contribute to overall few-shot classification performance?
\end{itemize}

\subsection{Datasets and Evaluation Methodology}

We evaluate \ourmethod\ on nine publicly available datasets across three domains: \textit{Application Identification}, \text{Device Identification}, and \textit{ Attack Identification}. All datasets are transformed into a multivariate time-series format. To ensure comparability across tasks and domains, we normalize the number of samples per class during task construction and discard classes with insufficient data. Importantly, each dataset is used for meta-training or evaluation, but never both, ensuring a strict assessment of generalization.

\begin{table*}[t]
\centering
\caption{Summary of Datasets Used in Evaluation}
\label{tab:datasets}
\begin{tabular}{l|c|c|c|c}
\hline
\textbf{Dataset} & \textbf{Task} & \textbf{\# Classes} & \textbf{Samples per Class} & \textbf{Protocols} \\
\hline
Appsniffer (No-VPN)~\cite{Appsniffer} & Application Identification & 150 & 50 & TCP, TLS \\
Appsniffer (SuperVPN)~\cite{Appsniffer} & Application Identification & 150 & 50 & UDP, VPN \\
Appsniffer (NordVPN)~\cite{Appsniffer} & Application Identification & 150 & 50 & UDP, VPN \\
Appsniffer (TurboVPN)~\cite{Appsniffer} & Application Identification & 150 & 50 & UDP, VPN \\
Appsniffer (Surfshark)~\cite{Appsniffer} & Application Identification & 150 & 50 & TCP, UDP, VPN \\
\hline
UNSW-IoT~\cite{unsw_IoT} & Device Identification & 27 & 1--20 & TCP, UDP, TLS \\
Aalto-IoT~\cite{aalto_IoT} & Device Identification & 28 & 5--20 & TCP, UDP \\
\hline
CIC-IDS 2017~\cite{cicids2017} & Attack Detection & 8 & 192--138{,}957 & TCP, UDP, TLS \\
ToN-IoT~\cite{ton_IoT} & Attack Detection & 9 & 12 & TCP, MQTT \\
\hline
\end{tabular}
\smallskip

\noindent\textit{Abbreviations:} \textbf{ANV} – Appsniffer (No-VPN), \textbf{ASV} – Appsniffer (SuperVPN), \textbf{ANVn} – Appsniffer (NordVPN), \textbf{ATV} – Appsniffer (TurboVPN), \textbf{ASh} – Appsniffer (Surfshark), \textbf{UIoT} – UNSW-IoT, \textbf{AIoT} – Aalto-IoT, \textbf{CIC17} – CIC-IDS 2017, \textbf{ToN} – ToN-IoT. These abbreviations will be used when referring to datasets in the plots.
\end{table*}

\textbf{Application Identification Datasets:}
To test \ourmethod 's performance on application classification, we use five datasets derived from the Appsniffer suite~\cite{Appsniffer}. Each dataset includes encrypted traffic from the same set of 150 Android mobile applications, captured under different network configurations: four commercial VPN services (SuperVPN, NordVPN, TurboVPN, Surfshark) and a No-VPN scenario. Each dataset consists of 7,500 samples (50 flows per class).

The VPN datasets encompass diverse encryption and tunneling protocols, including PPTP with RC4-based MPPE~\cite{ppp,mppc}, IKEv2/IPsec with ESP~\cite{ikev2,esp}, OpenVPN, and WireGuard~\cite{wireguard}. Each employs distinct key-exchange and encapsulation mechanisms, resulting in heterogeneous structural and temporal traffic characteristics. This diversity provides a strong basis for evaluating \ourmethod's ability to generalize across different encryption and tunneling conditions.

\textbf{Device Identification Datasets:} For IoT device classification, we use the UNSW-IoT and Aalto-IoT datasets, both of which contain encrypted traffic from smart home and mobile devices. UNSW-IoT~\cite{unsw_IoT} includes traffic from 27 device types, such as smart speakers, plugs, and hubs, with highly imbalanced class distributions; some devices have as few as a single sample. We retain only those with sufficient data, resulting in 20 device classes for evaluation. Aalto-IoT~\cite{aalto_IoT} includes traffic from 28 IoT devices, with more uniform sampling across classes (roughly 20 samples per device), ensuring balanced evaluation.

\textbf{Attack Detection Datasets:} To evaluate encrypted traffic classification in security scenarios, we use the CIC-IDS 2017~\cite{cicids2017} and TON-IoT~\cite{ton_IoT} datasets. CIC-IDS 2017~\cite{cicids2017} includes both benign and malicious traffic, covering various attack types including port scanning, DDoS, and brute-force attacks. While heavily imbalanced, it allows evaluation of the robustness of few-shot learning in realistic intrusion-detection scenarios. TON-IoT~\cite{ton_IoT} offers balanced samples across nine attack classes, each with 12 flows. The compact size and class diversity make it ideal for evaluating few-shot classification in constrained settings.

\textbf{\textit{Baseline Models:}}
We compare \ourmethod\ against four few-shot ETA baselines: MetaMRE~\cite{Yang2023}, RBRN~\cite{Zheng2020}, UMVD-FSL~\cite{UMVD}, and MeTaRocket~\cite{MetaRocket}. MetaMRE,MeTaRocket and RBRN use traffic metadata representations, UMVD-FSL operates on raw traffic flow bytes, and MeTaRocket likewise targets few-shot adaptation from sequence features. Evaluating \ourmethod\ against large-scale fully supervised models would be inappropriate, as such methods fall outside the scope of few-shot learning and do not address the data-scarcity challenges that GETA targets.

\textbf{\textit{Evaluation Methodology:}}
\label{sec:eval-protocol}
\label{sec:evaluation_methodology}
We evaluate all methods under a unified few-shot meta-learning protocol and report macro-F1 as the primary metric. For each scenario, we construct $N$-way, $K$-shot episodes with a disjoint query set, resampling support/query examples per episode, and preventing sample overlap within each episode. Intra-domain, cross-domain, and VPN settings follow the scenario definitions in Section~\ref{sec:evaluation}, with strict train/test dataset separation and identical transfer directions across methods. Unless otherwise stated, each scenario is evaluated on $n_{\mathrm{ep}}{=}128$ meta-test episodes. For episode $i$, macro-F1 is computed as
{\abovedisplayskip=0pt
\belowdisplayskip=0pt
\begin{align*}
\mathrm{F1}_{\mathrm{macro}}^{(i)} &= \frac{1}{C}\sum_{c=1}^{C}
\frac{2\,\mathrm{Precision}_c^{(i)}\,\mathrm{Recall}_c^{(i)}}
{\mathrm{Precision}_c^{(i)}+\mathrm{Recall}_c^{(i)}+\epsilon}, \\
\overline{\mathrm{F1}}_{\mathrm{macro}}^{(r)} &= \frac{1}{n_{\mathrm{ep}}}\sum_{i=1}^{n_{\mathrm{ep}}}\mathrm{F1}_{\mathrm{macro}}^{(i,r)},
\end{align*}
}%
where $C$ is the number of classes per episode, superscript $(r)$ indexes an independent training run with its own random seed, and $\epsilon$ is a small constant for numerical stability. We repeat the full protocol for $R=5$ seeds and report the grand mean $\overline{\mathrm{F1}}=\frac{1}{R}\sum_{r=1}^{R}\overline{\mathrm{F1}}_{\mathrm{macro}}^{(r)}$ with two-sided 95\% confidence intervals computed with Student's $t$ on these run-level aggregates,
\begin{align*}
\overline{\mathrm{F1}} \pm t_{0.975,\,R-1}\cdot \frac{s_{\mathrm{seed}}}{\sqrt{R}},
\end{align*}
where $s_{\mathrm{seed}}$ is the sample standard deviation of the $R$ values $\overline{\mathrm{F1}}_{\mathrm{macro}}^{(r)}$ (each already averaged over the $n_{\mathrm{ep}}$ episodes at the checkpoint chosen for run $r$). The same episode construction, $n_{\mathrm{ep}}$, checkpoint rule, and seed-level $t$ interval are used for GETA and all baselines (MetaRocket, MetaMRE, RBRN, and UMVD) whenever multi-seed reruns are reported, so differences in performance reflect model behavior rather than protocol mismatch. During meta-training, we select checkpoints by maximizing a running average of meta-test accuracy over the three most recent evaluation intervals, with each evaluation using the $n_{\mathrm{ep}}$ meta-test episodes defined above. Overall, we keep scenario definitions, $N/K$ settings, episode budgets, and metrics consistent across methods.

\subsection{GETA's Effectiveness Across RQs}
We now discuss the results for each research question, showing that GETA provides a robust and effective solution across all evaluation scenarios.

\noindent \textbf{\textit{RQ1: Generalization Across Classification Tasks}} 

A core challenge in encrypted traffic classification is achieving generalization beyond training domains. To evaluate this, we meta-train \ourmethod\ on one dataset and test it on distinct datasets under two settings:

\textbf{Intra-domain generalization:} Training and testing occur within the same task (e.g., application identification) but across different datasets or traffic conditions, such as VPN vs.\ non-VPN. This tests robustness to realistic variations within a domain.

\textbf{Cross-domain generalization:} Training is performed on one task (e.g., IoT device identification) and evaluation on another (e.g., application or attack classification), assessing whether \ourmethod\ learns transferable representations that remain effective for unseen domains with limited labeled data.

\textbf{Experimental Setup:}  
We evaluate intra- and cross-domain generalization on six public datasets spanning three domains: application identification, IoT device identification, and attack detection (two per domain). All experiments follow a 2-way 5-shot setup.  
For \textit{intra-domain generalization}, \ourmethod\ is trained and tested on the same task type but under different conditions (e.g., training on VPN traffic and testing on No-VPN traffic).  
For \textit{cross-domain generalization}, the model is trained on one domain (e.g., device identification) and evaluated on another (e.g., application or attack classification) to assess transferability across task semantics.

\begin{figure*}[!htbp]
\centering
\includegraphics[width=\textwidth]{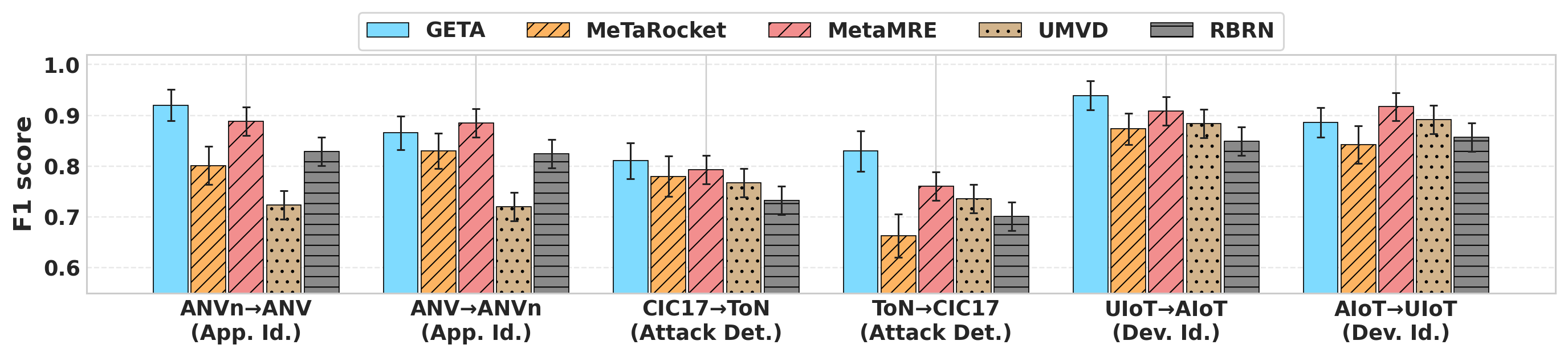}
\caption{Results for intra-domain tasks.}
\label{fig:intra_domain_accuracy}
\end{figure*}

\begin{figure*}[!htbp]
\centering
\includegraphics[width=\textwidth, trim=0cm 0cm 0cm 1.05cm, clip]{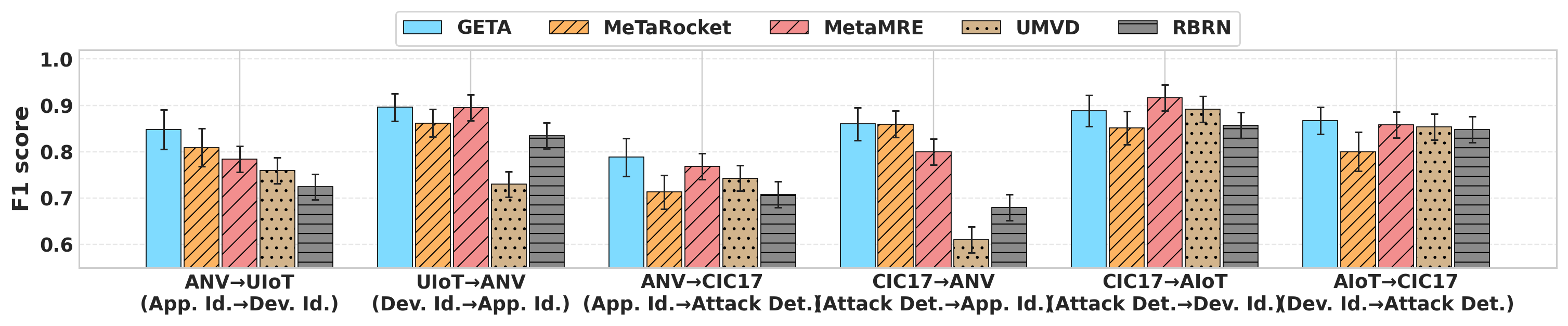}
\caption{Results for cross-domain transfer tasks.}
\label{fig:cross_domain_accuracy}
\vspace{-10pt}
\end{figure*}

\textbf{Results:}  
Figures~\ref{fig:intra_domain_accuracy} and~\ref{fig:cross_domain_accuracy} show that \ourmethod\ consistently outperforms all baselines in both settings.  
Intra-domain results confirm strong robustness to traffic variations, for instance, training on NordVPN and testing on No-VPN yields up to 8\% higher accuracy than competing models.  
Under cross-domain evaluation, baselines such as RBRN and MetaMRE fluctuate across datasets, while \ourmethod\ maintains stable accuracy, demonstrating its ability to learn domain-invariant representations and its effectiveness for real-world deployments with limited labeled data.

\noindent \textbf{\textit{RQ2: Impact of N-way K-shot Configurations}}  

We assess \ourmethod's robustness to task complexity by varying the number of classes ($N$) and labeled samples per class ($K$). This analysis examines how performance scales from simple binary to multi-class scenarios, reflecting realistic ETA deployments that must handle diverse traffic types.
\textbf{Experimental Setup:}  
We construct meta-training and evaluation tasks with varying $N$ and $K$ under strict cross-dataset conditions to test generalization. Specifically, training and testing datasets are disjoint: NordVPN $\rightarrow$ No-VPN (application), UNSW-IoT $\rightarrow$ Aalto-IoT (device), and ToN-IoT $\rightarrow$ CIC-IDS2017 (attack). This setup ensures evaluation on unseen domains that are representative of real-world conditions.

\FloatBarrier  % forces fig 1 to be placed before fig 2

\begin{figure*}[htbp]
\centering

\includegraphics[width=\textwidth, trim=0cm 0cm 0cm 2.2cm, clip]{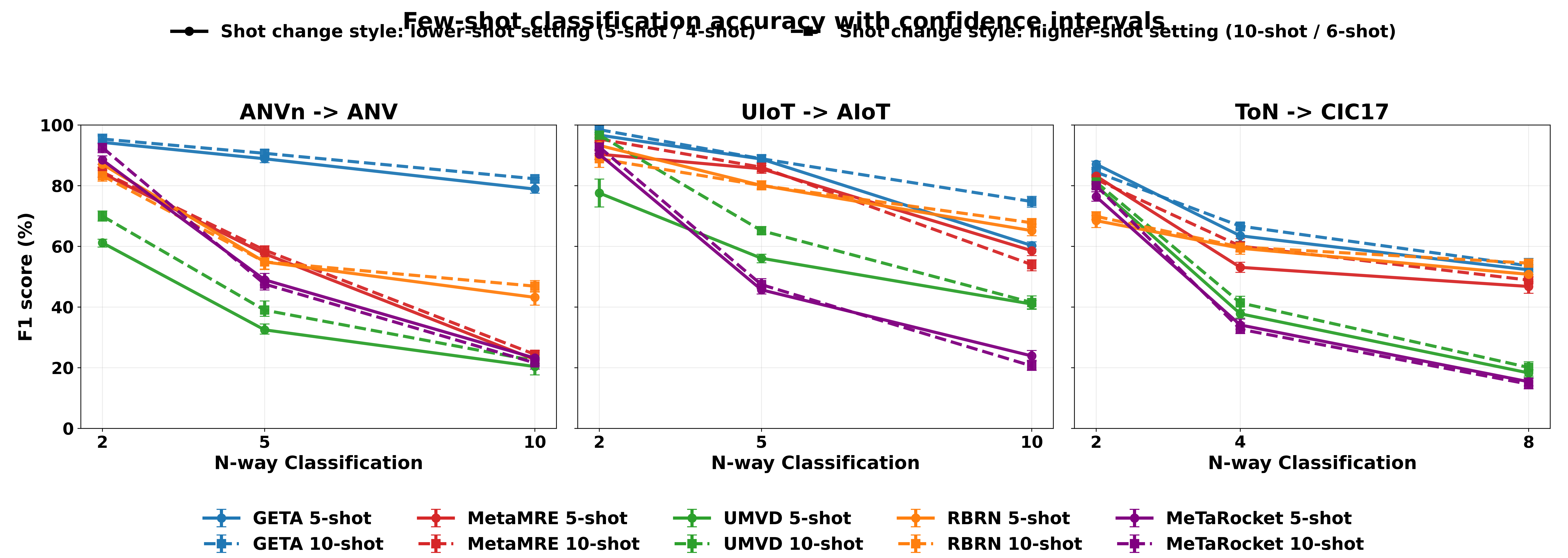}
\caption{Few-shot accuracy across N-way settings, with standard deviations. Solid and dashed lines denote lower and higher K-shot settings, respectively. GETA consistently outperforms the baselines.}
%\caption{Few-shot classification accuracy with standard deviations across different N-way settings. Line plots clearly show the degradation in performance as classification complexity increases. Solid lines represent lower, while dashed lines represent higher K-shot scenarios. GETA consistently outperforms across all datasets.}
\label{fig:fewshot_results}
\end{figure*}

\begin{figure*}[!htbp]

\centering
\includegraphics[width=\textwidth, trim=0cm 0.9cm 0cm 0cm, clip]{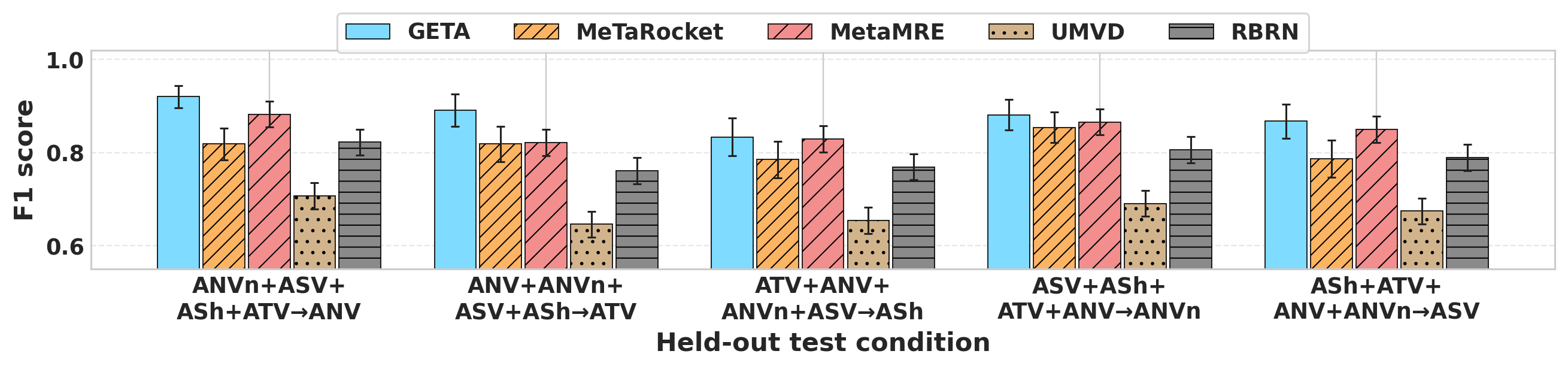}

\caption{Comparison of classification accuracy across VPN dataset combinations.}
\label{fig:vpn_classification_comparison}
\end{figure*}

\begin{table}[h]
    \centering
    \vspace{-5pt}
    \caption{Few-shot macro-F1 on NordVPN $\rightarrow$ No-VPN under high-$N$ settings (GETA only).}
    
    \label{tab:rq2_high_nway}
    \renewcommand{\arraystretch}{1.2}
    \setlength{\tabcolsep}{3.5pt}
    \begin{tabular}{c|c|c|c|c}
    % \toprule
      \hline
        \textbf{Train} & \textbf{Test} & \textbf{$N$-} & \textbf{$K$-} & \textbf{GETA} \\
        \textbf{Dataset} & \textbf{Dataset} & \textbf{way} & \textbf{Shots} & \textbf{F1 score} \\
              \hline
        % \midrule
        \multirow{6}{*}{NordVPN} & \multirow{6}{*}{No-VPN} 
            & 20 & 5 & \textbf{0.701 $\pm$ 0.021} \\ 
            & & & 10 & \textbf{0.773 $\pm$ 0.016} \\
            & & 50 & 5 & \textbf{0.622 $\pm$ 0.015} \\
            & & & 10 & \textbf{0.713 $\pm$ 0.011} \\
            & & 100 & 5 & \textbf{0.540 $\pm$ 0.017} \\
            & & & 10 & \textbf{0.647 $\pm$ 0.011} \\
        \bottomrule
    \end{tabular}
    \vspace{-10pt}
\end{table}

\textbf{Results:}  
As shown in Figure~\ref{fig:fewshot_results}, \ourmethod\ consistently achieves the highest accuracy across all N-way, K-shot settings. While baseline methods degrade sharply as the number of classes increases, \ourmethod\ degrades gracefully, with only a 12\% drop from 2-way 5-shot to 10-way 5-shot. Notably, it outperforms RBRN by up to 35\% in the 10-way 10-shot setting.

Extended evaluations (Table~\ref{tab:rq2_high_nway}) up to 100-way classification confirm stable performance, with minimal accuracy loss even in 50- and 100-way tasks, demonstrating scalability. Across IoT and attack detection tasks, \ourmethod\ remains the top-performing option in most configurations, with only minor degradation in selected high-complexity settings.  
Overall, these results show that \ourmethod\ offers strong cross-dataset generalization, stable performance, and practical scalability for real-world ETA.

\begin{table*}[ht]
\centering
\caption{Ablation study on the impact of major components in the proposed framework.}
\label{tab:ablation}
\small
\setlength{\tabcolsep}{3pt}
\begin{tabular}{lcccc|cc}
\toprule
\textbf{Variant} & \textbf{Base} & \textbf{Cls.\ head} & \textbf{Proto.\ enh.} & \textbf{MAML} & \textbf{Accuracy} & \textbf{F1} \\
\midrule
Full model & \checkmark & \checkmark & \checkmark & \checkmark & \underline{\textbf{0.921 $\pm$ 0.019}} & \underline{\textbf{0.921 $\pm$ 0.018}} \\
w/o class head & \checkmark & \xmark & \checkmark & \checkmark & \textbf{0.871 $\pm$ 0.011} & \textbf{0.870 $\pm$ 0.012} \\
w/o prototype enhancement & \checkmark & \checkmark & \xmark & \checkmark & 0.851 $\pm$ 0.019 & 0.850 $\pm$ 0.017 \\
w/o MAML & \checkmark & \checkmark & \checkmark & \xmark & 0.853 $\pm$ 0.014 & 0.852 $\pm$ 0.012 \\
\bottomrule
\end{tabular}
\vspace{-10pt}
\end{table*}
\noindent \textbf{\textit{RQ3: Effectiveness of \ourmethod\ in VPN Traffic}} 
Header-based models struggle in VPN environments where tunneling encapsulates original packet headers ~\cite{Appsniffer}. To evaluate robustness, we compare \ourmethod\ with three metadata-based baselines (MetaRocket, MetaMRE, RBRN) and one header-dependent model (UMVD).

\textbf{Experimental Setup:}  
We use five datasets~\cite{Appsniffer} that contain identical applications under both VPN and non-VPN conditions. We adopt a 2-way 5-shot setting with cross-validation training on four VPN datasets and testing on the remaining one, rotating across all combinations.

\textbf{Results:}  
As shown in Figure~\ref{fig:vpn_classification_comparison}, UMVD performs poorly under VPNs due to header obfuscation, while metadata-based methods remain stable. \ourmethod\ achieves the highest accuracy across all VPN settings, outperforming MetaMRE and RBRN by up to 5\%, demonstrating strong robustness when header information is unavailable.

\noindent \textbf{\textit{RQ4: Contribution of GETA's Components and Hyperparameters}}  
We analyze the impact of two core components: the prototype enhancement (embedding enhancer and self-attention) and the class-specific head, alongside the effect of MAML through ablation studies. We also assess GETA's sensitivity to two key hyperparameters: the $\lambda$ weighting value and the input packet sequence length.

\textbf{Experimental Setup:}  

We remove each component in turn and evaluate on few‑shot tasks meta‑trained on NordVPN and tested on No‑VPN. All variants share the same Transformer encoder for fair comparison. For hyperparameter analysis, we vary $\lambda$ from 0.0 to 1.0 and test packet sequence lengths ranging from 5 to 1,000.

\textbf{Results:}  
Table~\ref{tab:ablation} demonstrates that removing prototype enhancement causes the largest performance drop (~7\%), confirming its importance in refining support embeddings. Excluding the class-specific head reduces accuracy by ~5\%, further validating both modules as important to GETA’s few-shot capability. 

\begin{figure}[ht]
    \vspace{-10pt}
    \centering
    \includegraphics[width=\columnwidth,trim=0cm 0cm 0cm 0.9cm, clip]{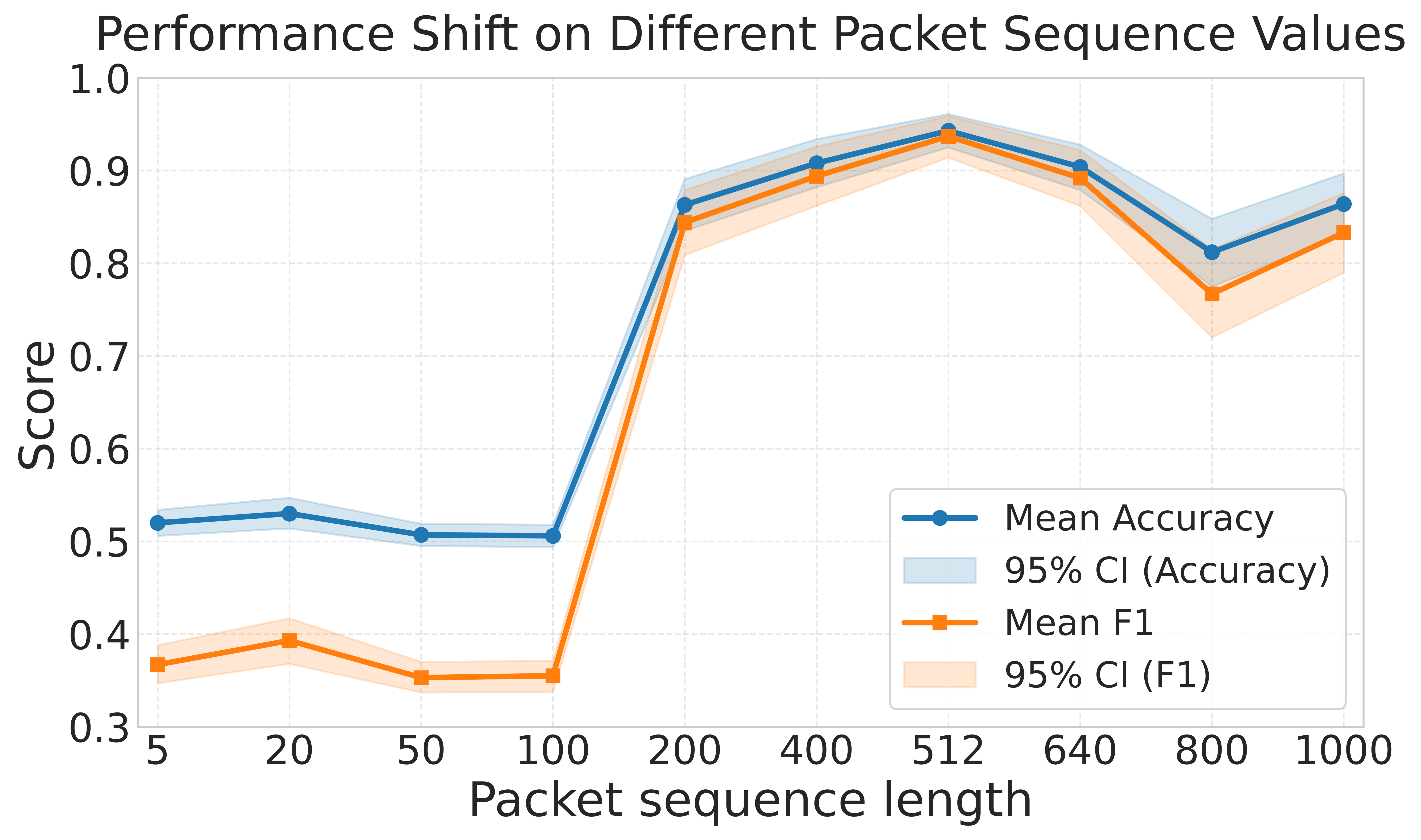}
    \caption{Performance shift across different packet sequence lengths.}
    \label{fig:packet_sequence}
\end{figure}

Table~\ref{tab:lambda_ablation} indicates that $\lambda = 0.5$ provides the best balance for the dual-pathway loss. As shown in Fig.~\ref{fig:packet_sequence}, performance is also highly sensitive to packet sequence length: accuracy is low for sequences of 100 packets or fewer, rises sharply at 200 packets, and peaks at 512 packets, which provides sufficient temporal context without adding excessive network noise.

\begin{table}[ht]
\centering
\caption{F1 score across different $\lambda$ values}
\label{tab:lambda_ablation}
\small
\setlength{\tabcolsep}{8pt}
\begin{tabular}{lcccccc}
\toprule
\textbf{$\lambda$} & 0.0 & 0.2 & 0.5 & 0.6 & 0.8 & 1.0 \\
\midrule
\textbf{F1} & 0.870 & 0.902 & 0.920 & 0.914 & 0.889 & 0.850 \\
\bottomrule
\end{tabular}
\vspace{-5pt}
\end{table}

\vspace{-4 pt}
\section{Conclusion}
\vspace{-4 pt}
This work presented GETA, a meta-learning-based framework for ETA that integrates a state-of-the-art time-series model with enhanced embedding techniques incorporating packet timing, packet size, and direction. GETA consistently outperformed state-of-the-art traffic metadata-based techniques, particularly in challenging scenarios such as VPN-encrypted environments. It further demonstrated strong adaptability in few-shot conditions, achieving high accuracy with only a few traffic flows per class, an important capability for scenarios where labeled data is scarce. By relying solely on traffic metadata, GETA effectively generalizes across various network conditions, as validated experimentally, positioning it as a promising solution for scalable cross-domain ETA.

% \section{Summary of Findings}

% Our evaluation highlights three key insights:
% \begin{itemize}
%     \item Metadata-based classification is more effective than header-based models, particularly in VPN environments where header information is obfuscated.
%     \item Our model generalizes well across application, device, and attack classification tasks, demonstrating robustness in diverse encrypted traffic datasets.
%     \item Few-shot learning improves classification performance with minimal labeled data, making it a practical solution for real-world deployment where large-scale labeling is infeasible.
% \end{itemize}

% These findings underscore the advantages of metadata-driven traffic analysis and adaptive learning techniques for encrypted traffic classification in dynamic network environments.

% \section*{Acknowledgment}

% The preferred spelling of the word ``acknowledgment'' in America is without 
% an ``e'' after the ``g''. Avoid the stilted expression ``one of us (R. B. 
% G.) thanks $\ldots$''. Instead, try ``R. B. G. thanks$\ldots$''. Put sponsor 
% acknowledgments in the unnumbered footnote on the first page.

\bibliographystyle{IEEEtran}
\bibliography{bibliography}

% Add to preamble: \usepackage{afterpage}

\clearpage
\appendix

\section{Discussion}

\textbf{Key Findings:}  
\ourmethod\ demonstrates robust performance across diverse encrypted traffic classification tasks, outperforming metadata-based baselines (MetaMRE, RBRN) and header-dependent methods (UMVD). Its incorporation of inter-arrival time alongside packet size and direction enriches temporal context, enabling superior generalization compared to methods that omit timing information.

% \textbf{Domain-Matched Noise Regularization:}  
% A key insight from the NordVPN and No-VPN datasets (Figure~\ref{fig:nordvpn_novpn_patterns}) reveals the importance of training data noise characteristics. NordVPN flows exhibit greater bidirectional activity and packet size variance, creating a noisier temporal structure. When meta-trained on such noisy data, \ourmethod\ learns domain-invariant features through what we term \emph{domain-matched noise regularization}, yielding:
% \begin{itemize}[after=\vspace{-0.67em}]
%     \item \textbf{NordVPN $\rightarrow$ No-VPN:} Strong transfer from noisy to clean domains.  
%     \item \textbf{No-VPN $\rightarrow$ NordVPN:} F1 score drops when clean-trained models face unseen noise patterns.  
%     \item \textbf{IoT/Attack $\rightarrow$ No-VPN:} Domain-matched noise can outweigh benefits from broader training diversity.  
% \end{itemize}
% This phenomenon enables reliable inference across unfamiliar networks with minimal labeled data.

\begin{figure}[htbp]
    \vspace{-10 pt}
    \centering
    \includegraphics[width=\columnwidth, trim=0cm 13cm 0cm 20.5cm, clip]{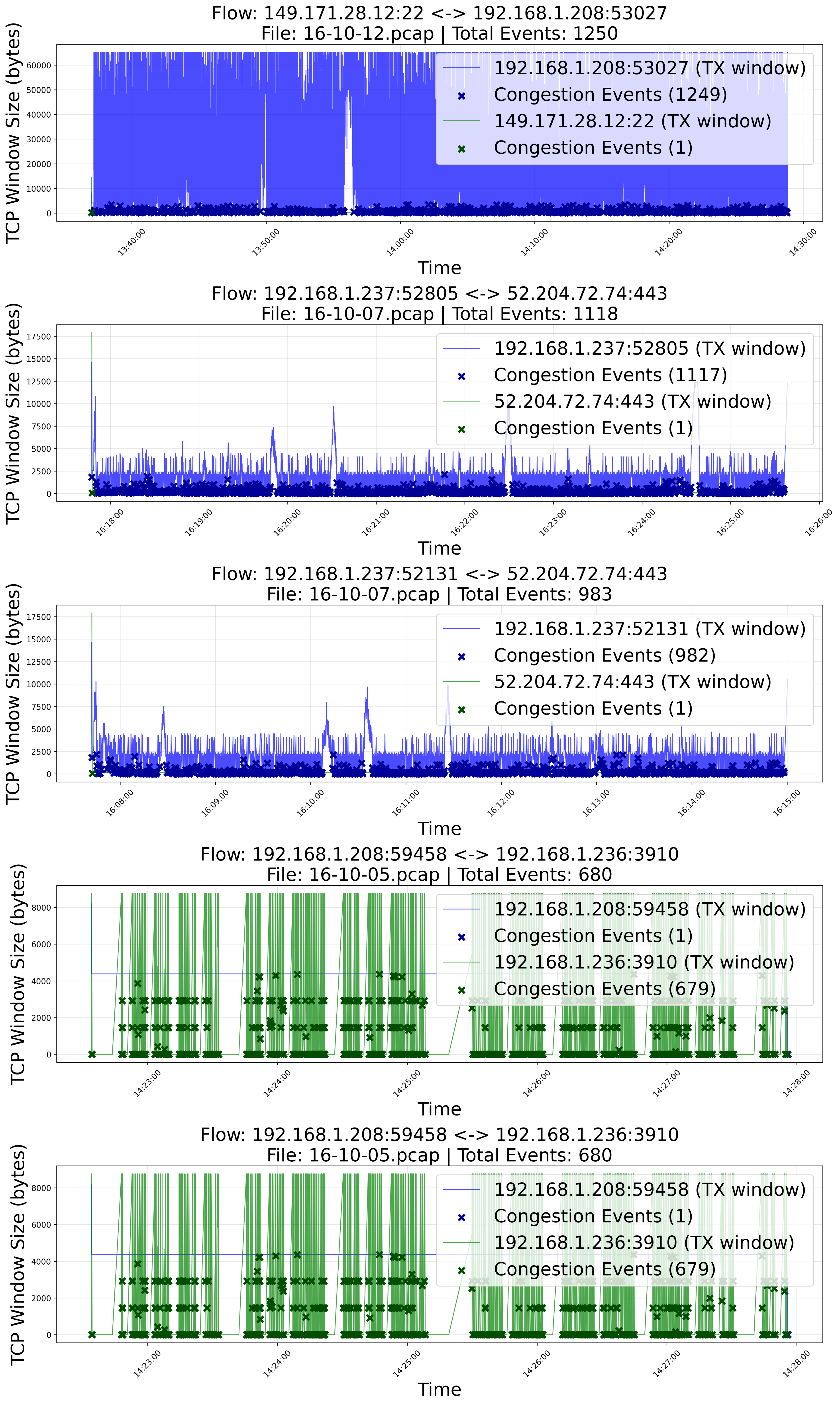}
    \caption{Four of the five datasets contain congestion and retransmission patterns characteristic of TCP traffic. This figure illustrates congestion behavior in the UNSW-IoT dataset, showing drops in TCP window size that indicate congestion events and reflect network-level congestion.}
    \label{fig:tcp_window_drop}
    \vspace{-10pt}
\end{figure}

\textbf{Robustness to Network Variability:}  
 Metadata features such as packet size, timing, and direction can be distorted by MTU changes, congestion, retransmissions, or protocol behavior (see Figure~\ref{fig:tcp_window_drop}). Despite this, \ourmethod\ remains robust by modeling aggregate flow behavior rather than brittle packet-level bytes. This makes it resilient to encryption and network noise, maintaining visibility under adverse conditions

\textbf{Scalability and Efficiency:}
\begin{figure}[!t]
\centering
\begin{tikzpicture}
\begin{axis}[
    width=7cm,  % Reduced from 12cm
    height=5cm,   % Reduced from 6cm
    xlabel={Task Configuration},
    ylabel={Memory Allocated (MB)},
    title={Resource Consumption vs Task Complexity},
    symbolic x coords={2-way\\5-shot, 2-way\\10-shot, 5-way\\5-shot, 5-way\\10-shot, 10-way\\5-shot, 10-way\\10-shot},
    xtick=data,
    x tick label style={
        rotate=0, 
        anchor=north,
        font=\small,
        align=center,
        text width=1.2cm,
        yshift=-2pt,
    },
    legend style={at={(0.98,0.10)}, anchor=south east, legend columns=1}, 
    grid=major,
    grid style={dashed,gray!30},
    axis y line*=left,
    axis x line*=bottom,
    ymin=0,
    ymax=900,
    enlarge x limits=0.1,
]
\addplot[thick, blue, mark=*, mark size=3pt] coordinates {
    (2-way\\5-shot, 164.802)
    (2-way\\10-shot, 228.794)
    (5-way\\5-shot, 293.944)
    (5-way\\10-shot, 453.979)
    (10-way\\5-shot, 504.247)
    (10-way\\10-shot, 824.410)
};
\addlegendentry{Memory (MB)}
\end{axis}
\begin{axis}[
    width=7cm,  % Reduced from 12cm
    height=5cm,   % Reduced from 6cm
    symbolic x coords={2-way\\5-shot, 2-way\\10-shot, 5-way\\5-shot, 5-way\\10-shot, 10-way\\5-shot, 10-way\\10-shot},
    xtick=data,
    x tick label style={
        rotate=0, 
        anchor=center,
        font=\small,
        align=center,
        text width=1.2cm,
    },
    legend style={at={(0.10,0.90)}, anchor=north west, legend columns=1},
    axis y line*=right,
    axis x line=none,
    ylabel={Inference Time (ms)},
    ylabel near ticks,
    ylabel style={rotate=180},
    ymin=6,
    ymax=16,
    enlarge x limits=0.1,
]
\addplot[thick, red, mark=square*, mark size=3pt] coordinates {
    (2-way\\5-shot, 8.655)
    (2-way\\10-shot, 9.128)
    (5-way\\5-shot, 11.322)
    (5-way\\10-shot, 11.490)
    (10-way\\5-shot, 13.622)
    (10-way\\10-shot, 14.239)
};
\addlegendentry{Inference Time (ms)}
\end{axis}
\end{tikzpicture}
\caption{Memory allocation and inference time across different task configurations}
\label{fig:resource_consumption}
\vspace{-15pt}
\end{figure}
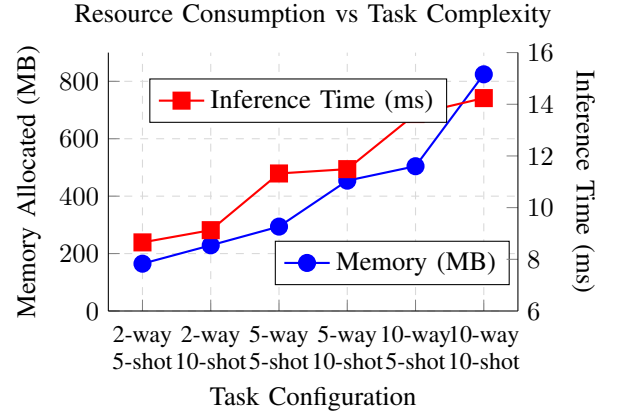

As demonstrated in RQ2, \ourmethod\ exhibits controlled performance degradation as task complexity increases, unlike baselines that collapse sharply in high N-way settings. Resource consumption (Figure~\ref{fig:resource_consumption}) remains manageable, with memory scaling from 165MB to 824MB and latency from 8.7ms to 14.2ms across increasing task complexity.

\subsection{Limitations}
While our proposed method demonstrates robust adaptability and high classification accuracy across diverse encrypted network conditions, it fundamentally relies on a few-shot learning paradigm. Specifically, the model requires a small, labeled support set for each novel traffic class to successfully fine-tune its decision boundaries. In highly dynamic real-world environments where emerging applications, novel IoT devices, or zero-day threats appear with absolutely no prior labeled samples, this dependency becomes a bottleneck. The current architecture cannot autonomously classify completely unseen traffic classes without this minimal baseline of prior knowledge.

\subsection{Future Work}
To address the data-scarcity bottleneck, future research should explore Zero-Shot Learning (ZSL) to identify entirely unseen applications. However, transitioning to ZSL in encrypted environments is exceptionally challenging. Applying representation learning directly to raw bytes causes models to exploit spurious dataset shortcuts rather than learning true protocol semantics, leading to catastrophic generalization failures \cite{wickramasinghe2025sok}. Furthermore, while utilizing protocol-agnostic metadata such as our multivariate time-series features avoids these raw-byte shortcuts, it remains unclear how metadata alone can effectively translate to generalized zero-shot classification. A fundamental semantic gap exists because the feature space of encrypted traffic is notoriously high-dimensional and noisy, making it difficult to optimize a well-structured zero-shot embedding space without strong prior knowledge. Therefore, future ZSL implementations must investigate how to mathematically fuse robust metadata representations with external semantic knowledge (e.g., textual application descriptions or relational graphs) to accurately bridge this inferential gap.

\end{document}